\documentclass[12pt]{article}
\usepackage{graphicx}
\usepackage{lscape}
\usepackage{citesort}
\usepackage{amssymb}
\usepackage{appendix}
\usepackage{multirow}

\begin{document}
\def\qq{\langle \bar q q \rangle}
\def\uu{\langle \bar u u \rangle}
\def\dd{\langle \bar d d \rangle}
\def\sp{\langle \bar s s \rangle}
\def\GG{\langle g_s^2 G^2 \rangle}
\def\Tr{\mbox{Tr}}
\def\figt#1#2#3{
        \begin{figure}
        $\left. \right.$
        \vspace*{-2cm}
        \begin{center}
        \includegraphics[width=10cm]{#1}
        \end{center}
        \vspace*{-0.2cm}
        \caption{#3}
        \label{#2}
        \end{figure}
	}
	
\def\figb#1#2#3{
        \begin{figure}
        $\left. \right.$
        \vspace*{-1cm}
        \begin{center}
        \includegraphics[width=10cm]{#1}
        \end{center}
        \vspace*{-0.2cm}
        \caption{#3}
        \label{#2}
        \end{figure}
                }

\def\ds{\displaystyle}
\def\beq{\begin{equation}}
\def\eeq{\end{equation}}
\def\bea{\begin{eqnarray}}
\def\eea{\end{eqnarray}}
\def\beeq{\begin{eqnarray}}
\def\eeeq{\end{eqnarray}}
\def\ve{\vert}
\def\vel{\left|}
\def\ver{\right|}
\def\nnb{\nonumber}
\def\ga{\left(}
\def\dr{\right)}
\def\aga{\left\{}
\def\adr{\right\}}
\def\lla{\left<}
\def\rra{\right>}
\def\rar{\rightarrow}
\def\lrar{\leftrightarrow}  
\def\nnb{\nonumber}
\def\la{\langle}
\def\ra{\rangle}
\def\ba{\begin{array}}
\def\ea{\end{array}}
\def\tr{\mbox{Tr}}
\def\ssp{{\Sigma^{*+}}}
\def\sso{{\Sigma^{*0}}}
\def\ssm{{\Sigma^{*-}}}
\def\xis0{{\Xi^{*0}}}
\def\xism{{\Xi^{*-}}}
\def\qs{\la \bar s s \ra}
\def\qu{\la \bar u u \ra}
\def\qd{\la \bar d d \ra}
\def\qq{\la \bar q q \ra}
\def\gGgG{\la g^2 G^2 \ra}
\def\q{\gamma_5 \not\!q}
\def\x{\gamma_5 \not\!x}
\def\g5{\gamma_5}
\def\sb{S_Q^{cf}}
\def\sd{S_d^{be}}
\def\su{S_u^{ad}}
\def\sbp{{S}_Q^{'cf}}
\def\sdp{{S}_d^{'be}}
\def\sup{{S}_u^{'ad}}
\def\ssp{{S}_s^{'??}}

\def\sig{\sigma_{\mu \nu} \gamma_5 p^\mu q^\nu}
\def\fo{f_0(\frac{s_0}{M^2})}
\def\ffi{f_1(\frac{s_0}{M^2})}
\def\fii{f_2(\frac{s_0}{M^2})}
\def\O{{\cal O}}
\def\sl{{\Sigma^0 \Lambda}}
\def\es{\!\!\! &=& \!\!\!}
\def\ap{\!\!\! &\approx& \!\!\!}
\def\ar{&+& \!\!\!}
\def\ek{&-& \!\!\!}
\def\kek{\!\!\!&-& \!\!\!}
\def\cp{&\times& \!\!\!}
\def\se{\!\!\! &\simeq& \!\!\!}
\def\eqv{&\equiv& \!\!\!}
\def\kpm{&\pm& \!\!\!}
\def\kmp{&\mp& \!\!\!}
\def\mcdot{\!\cdot\!}
\def\erar{&\rightarrow&}


\def\simlt{\stackrel{<}{{}_\sim}}
\def\simgt{\stackrel{>}{{}_\sim}}


\renewcommand{\textfraction}{0.2}    
\renewcommand{\topfraction}{0.8}   

\renewcommand{\bottomfraction}{0.4}   
\renewcommand{\floatpagefraction}{0.8}
\newcommand\mysection{\setcounter{equation}{0}\section}

\def\baeq{\begin{appeq}}     \def\eaeq{\end{appeq}}  
\def\baeeq{\begin{appeeq}}   \def\eaeeq{\end{appeeq}}
\newenvironment{appeq}{\beq}{\eeq}   
\newenvironment{appeeq}{\beeq}{\eeeq}
\def\bAPP#1#2{
 \markright{APPENDIX #1}
 \addcontentsline{toc}{section}{Appendix #1: #2}
 \medskip
 \medskip
 \begin{center}      {\bf\LARGE Appendix #1 :}{\quad\Large\bf #2}
\end{center}
 \renewcommand{\thesection}{#1.\arabic{section}}
\setcounter{equation}{0}
        \renewcommand{\thehran}{#1.\arabic{hran}}
\renewenvironment{appeq}
  {  \renewcommand{\theequation}{#1.\arabic{equation}}
     \beq
  }{\eeq}
\renewenvironment{appeeq}
  {  \renewcommand{\theequation}{#1.\arabic{equation}}
     \beeq
  }{\eeeq}
\nopagebreak \noindent}

\def\eAPP{\renewcommand{\thehran}{\thesection.\arabic{hran}}}

\renewcommand{\theequation}{\arabic{equation}}
\newcounter{hran}
\renewcommand{\thehran}{\thesection.\arabic{hran}}

\def\bmini{\setcounter{hran}{\value{equation}}
\refstepcounter{hran}\setcounter{equation}{0}
\renewcommand{\theequation}{\thehran\alph{equation}}\begin{eqnarray}}
\def\bminiG#1{\setcounter{hran}{\value{equation}}
\refstepcounter{hran}\setcounter{equation}{-1}
\renewcommand{\theequation}{\thehran\alph{equation}}
\refstepcounter{equation}\label{#1}\begin{eqnarray}}


\newskip\humongous \humongous=0pt plus 1000pt minus 1000pt
\def\caja{\mathsurround=0pt}


\title{
         {\Large
                 {\bf
Mixing angle of doubly heavy baryons in QCD
                 }
         }
      }

\author{\vspace{1cm}\\
{\small T. M. Aliev \thanks {e-mail:
taliev@metu.edu.tr}~\footnote{permanent address:Institute of
Physics,Baku,Azerbaijan}\,\,, K. Azizi \thanks {e-mail:
kazizi@dogus.edu.tr}\,\,, M. Savc{\i} \thanks
{e-mail: savci@metu.edu.tr}} \\
{\small Physics Department, Middle East Technical University,
06531 Ankara, Turkey }\\
{\small$^\ddag$ Physics Department,  Faculty of Arts and Sciences,
Do\u gu\c s University,} \\
{\small Ac{\i}badem-Kad{\i}k\"oy,  34722 Istanbul, Turkey}}

\date{}

\begin{titlepage}
\maketitle
\thispagestyle{empty}

\begin{abstract}

We calculate the mixing angles between the spin--1/2, $\Xi_{bc}$--$\Xi^\prime_{bc}$ and 
$\Omega_{bc}$--$\Omega^\prime_{bc}$ states of doubly heavy baryons
within the QCD sum rules method. It is found that the mixing angles are large
and have the values $\varphi_{\Xi_{bc}} = 16^0 \pm 5^0$ and
$\varphi_{\Omega_{bc}} = 18^0 \pm 6^0$, respectively. The mixing angles are slightly 
smaller compared to the predictions of the non--relativistic quark model,
$\varphi_{\Xi_{bc}} = 25.5^0$ and $\varphi_{\Omega_{bc}} = 25.9^0$.
\end{abstract}

~~~PACS numbers: 11.55.Hx, 14.20.--c, 14.20.Mr,  14.20.Lq
\end{titlepage}

\section{Introduction}

Baryons with two heavy quarks have been the subject of intensive theoretical
studies. The study of these baryons can provide useful information for
understanding the non--perturbative QCD effects. On the experimental
side, only one $\Xi_{cc}^{++}$ state is observed by the SELEX
Collaboration. However, the quark model predicts the existence of other
doubly heavy baryons, and their masses are estimated in this model (for a
review on doubly heavy baryons, see for instance \cite{Rdhb01}).   

Doubly heavy baryons also represent a very suitable framework for studying
the consequences of heavy quark spin symmetry \cite{Rdhb02}. According to
this symmetry, in the infinite heavy quark mass limit, the diquarks formed of
two heavy quarks can possess total spin $s=0$ or $1$. Taking into account the
spin of the light quark, the ground states of doubly heavy baryons can have
total spin of 1/2 or 3/2.

Since the heavy quark mass is finite, the hyperfine interaction between one
of the heavy quarks and light quark admix spin--0 and spin--1 components.
Obviously, this mixing for the baryons with two identical heavy quarks
should be very small, since the antisymmetry of the wave functions require
radial or higher orbital angular momentum states. But for the heavy baryons
with two different heavy quarks this mixing can be large in principle.
It is shown in \cite{Rdhb03} that the hyperfine mixing can considerably 
change the decay widths of doubly heavy baryons. The mixing problem of
doubly heavy baryons in semileptonic decays are discussed in many works
\cite{Rdhb04,Rdhb05,Rdhb06,Rdhb07}.

As has  been noted, the hyperfine mixing among the ground states of
the doubly heavy baryons is studied in \cite{Rdhb03} within the framework of
the quark model. The effects of this mixing for the electromagnetic decays
of the doubly heavy baryons are investigated in \cite{Rdhb08}. Calculation
of the mixing angle of baryons containing only one heavy quark within the
QCD sum rules method \cite{Rdhb09} is given in \cite{Rdhb10}.

In the present work, we generalize our previous study to the baryons
containing double heavy quarks, i.e., we calculate the mixing angle between   
$\Xi_{bc}$--$\Xi^\prime_{bc}$ and $\Omega_{bc}$--$\Omega^\prime_{bc}$ states
within the QCD sum rules approach.

\section{Mixing Angles Between the  $\Xi_{bc}$--$\Xi^\prime_{bc}$
and $\Omega_{bc}$--$\Omega^\prime_{bc}$ States}

In order to calculate the mixing angles between $\Xi_{bc}$--$\Xi^\prime_{bc}$
and $\Omega_{bc}$--$\Omega^\prime_{bc}$ states within the QCD sum rules
method, we consider the following correlation function:
\bea
\label{edhb01}
\Pi = i \int d^4x e^{iqx} \lla 0 \vel T\left\{ \eta_1 (x) \bar{\eta}_2 (0)
\right\} \ver 0 \rra~,
\eea
where $\eta_1$ and $\eta_2$ are the interpolating currents corresponding to
the physical states. Obviously, these currents should be linear combinations of
the interpolating currents of unmixed states $\eta_1^0$ and $\eta_2^0$, i.e.,
\bea
\label{edhb02}
\eta_1 \es   \cos(\varphi) \eta_1^0 + \sin(\varphi) \eta_2^0~, \nnb \\
\eta_2 \es - \sin(\varphi) \eta_1^0 + \cos(\varphi) \eta_2^0~,
\eea
 According to the sum rules philosophy, the correlation
function is calculated in two different ways, either in terms of hadronic 
parameters or
quark--gluon degrees of freedom. Once this is accomplished, these two
representations of the correlation function are equated, as a result of
which we obtain the QCD sum rules for the corresponding physical quantities.

When we saturate the correlation function given in Eq. (\ref{edhb01}) with
hadronic states we separate the ground state contributions, which should be equal
to zero since the physical ground states described by the
interpolating currents  $\eta_1$ and $\eta_2$ are orthogonal.
Here we would like to make the following two remarks. The mixing angles for
the excited states are generally different from that of the ground states.
For this reason, the physical part of the correlation function can get
non--zero contributions from excited and continuum states. However, in the sum
rules method, Borel transformation is performed in order to enhance the ground
state contribution (see below). After this transformation, the contributions
of the excited and continuum states are exponentially suppressed. Therefore,
non--vanishing contributions to the physical part of the correlation function
from the excited and continuum states should be very small.

Our second remark is related to the negative-parity baryon contributions to
the correlation function. In principle, the  negative parity baryons can give
contributions to the correlation function. These contributions disappear only
if their mixing angles  are the same as the one in the
interpolating current. We assume that this is the case here, so we neglect  the negative-parity
baryon contributions   in the present study.

Substituting Eq. (\ref{edhb02}) in Eq. (\ref{edhb01}) we get,
\bea
\label{edhb03}
\tan (2 \varphi) = {2 \Pi_{12}^{(0)} \over \Pi_{11}^{(0)} -\Pi_{22}^{(0)}}~,
\eea
where $\Pi_{ij}^{(0)}$ correspond to the correlation function,
\bea
\label{nolabel}
\Pi_{ij}^{(0)} = i \int d^4x e^{iqx} \lla 0 \vel T\left\{ \eta_i^{(0)} (x) 
\bar{\eta}_j^{(0)} (0) \right\} \ver 0 \rra~. \nnb
\eea

For interpolating currents $\eta_1^0$ and $\eta_2^0$ which correspond to the
unmixed states, we choose,
\bea          
\label{edhb04}
\eta_1^0 \es  {1\over \sqrt{2}} \epsilon^{abc} \Big\{ (b^{aT} C q^b)
\gamma_5 c^c + (c^{aT} C q^b) \gamma_5 b^c + t  (b^{cT} C \gamma_5 q^b) c^c
+ t  (c^{aT} C \gamma_5 q^b) b^c \Big\}~, \\ \nnb \\
\label{edhb05}
\eta_2^0 \es {1\over \sqrt{6}} \epsilon^{abc} \Big\{ 2 (b^{aT} C c^b)
\gamma_5 q^c + (b^{aT} C q^b) \gamma_5 c^c - (c^{aT} C q^b) \gamma_5 b^c +
2t  (b^{aT} C \gamma_5 c^b) q^c \nnb \\
\ar t  (b^{aT} C \gamma_5 q^b) c^c -
t  (c^{aT} C \gamma_5 q^b) b^c \Big\}~.
\eea
Considering the Lorentz invariance,  the two--point correlation function can
be written as:
\bea
\label{edhb06}
\Pi_{ij}^{(0)} = \Pi_{ij}^{(1)} (q^2) \, \rlap/{q} + \Pi_{ij}^{(2)} (q^2)
\, \mbox{I}~.
\eea
In further analysis of the mixing angle between the doubly heavy baryon
states, we shall take into consideration both $\rlap/{q}$ and I
structures.

The invariant functions $\Pi_{ij}^{(1)}$ and $\Pi_{ij}^{(2)}$ can be related
to their imaginary part with the help of the dispersion relation,
\bea
\label{edhb07} 
\Pi_{ij}^{(\alpha)} = \int_{(m_1+m_2)^2}^\infty {\rho_{ij}^{(\alpha)} (s) ds
\over s-q^2}~,
\eea
where $m_1$ and $m_2$ are heavy quarks masses and $\rho_{ij}^{(\alpha)}$ are the spectral densities which are given as:
\bea
\label{edhb08}
\rho_{ij}^{(\alpha)}(s) = {1\over \pi} \mbox{Im} \Pi_{ij}^{(\alpha)\,OPE}(s)~,
\eea
with the superscripts $\alpha=1$ and $2$ correspond to the structures $\rlap/{q}$
and I, respectively. The expressions for the spectral densities are
obtained as (see also  \cite{Rdhb11}):
\bea
\label{edhb09}
\rho_{11}^{(1)} \es {3\over 256 \pi^4} \int_{\alpha_{min}}^{\alpha_{max}}
{d\alpha \over \alpha} \int_{\beta_{min}}^{1-\alpha} {d\beta \over \beta}
(m_1^2 \beta + m_2^2 \alpha - s \alpha \beta) \nnb \\
\cp \Big\{ (m_1^2 \beta + m_2^2 \alpha - s \alpha \beta) (5 + 2 t + 5 t^2) -
2 (1-\alpha -\beta) (1-t)^2 m_1 m_2 \nnb \\
\ar 6 (1-t^2) m_q (m_1 \beta + m_2\alpha) \Big\} + 
{\qq \over 32 \pi^2}  \int_{\alpha_{min}}^{\alpha_{max}} d\alpha
\Big\{3 (1-t^2) \Big[ (1-\alpha) m_1 + \alpha m_2 \Big] \nnb \\
\ar m_q (1-\alpha)\alpha (5 + 2 t + 5 t^2) \Big\}~, \\ \nnb \\
\label{edhb10}
\rho_{12}^{(1)} \es {\sqrt{3} \over 64 \pi^4}
\int_{\alpha_{min}}^{\alpha_{max}}
{d\alpha \over \alpha} \int_{\beta_{min}}^{1-\alpha} {d\beta \over \beta}
(m_1^2 \beta + m_2^2 \alpha - s \alpha \beta) (-2 + t + t^2) m_q 
(\beta m_1 - \alpha m_2) \nnb \\
\ar {\qq \over 16 \sqrt{3} \pi^2} \int_{\alpha_{min}}^{\alpha_{max}} d\alpha
(-2 + t + t^2) \Big[ (1-\alpha) m_1 - \alpha m_2 \Big], \\ \nnb \\
\label{edhb11}
\rho_{22}^{(1)} \es {1 \over 256 \pi^4}
\int_{\alpha_{min}}^{\alpha_{max}} {d\alpha \over \alpha} 
\int_{\beta_{min}}^{1-\alpha} {d\beta \over \beta}
(m_1^2 \beta + m_2^2 \alpha - s \alpha \beta) \nnb \\
\cp \Big\{ 3 (m_1^2 \beta + m_2^2 \alpha - s \alpha \beta)(5 + 2 t + 5 t^2)
+ 2 (1-t) \Big[ (1-\alpha-\beta) (13 + 11 t) m_1 m_2 \nnb \\
\ar (1+ 5 t) m_q (\beta m_1 + \alpha m_2) \Big] \Big\} +
{\qq \over 96 \pi^2} \int_{\alpha_{min}}^{\alpha_{max}} d\alpha
\Big\{ 3 m_q (1-\alpha) \alpha ((5 + 2 t + 5 t^2) \nnb \\
\ar (1-t) (1+5 t) \Big[(1-\alpha) m_1 + \alpha m_2 \Big] \Big\}~, \\ \nnb \\
\label{edhb12}
\rho_{11}^{(2)} \es {3\over 256 \pi^4} \int_{\alpha_{min}}^{\alpha_{max}}
{d\alpha \over \alpha} \int_{\beta_{min}}^{1-\alpha} {d\beta \over \beta} 
(m_1^2 \beta + m_2^2 \alpha - s \alpha \beta) \nnb \\
\cp \Big\{ -3 \alpha (1-t^2) {m_2\over \beta} (m_1^2 \beta + m_2^2 \alpha -
s \alpha \beta) + m_1 \Big[ -3 \beta (1-t^2) {1\over \alpha \beta} (m_2^2
\beta + m_2^2 \alpha - s \alpha \beta) \nnb \\
\ek 2 m_1 m_2 (5 + 2 t + 5 t^2) \Big]
\Big\} + {\qq\over 64 \pi^2} \int_{\alpha_{min}}^{\alpha_{max}} d\alpha
\Bigg\{(1-\alpha) \alpha (1-t)^2 \Bigg[ 3 m_0^2 \nnb \\
\ar {4 [m_1^2 (1-\alpha) + m_2^2 \alpha - s
\alpha (1-\alpha)]\over \alpha (1-\alpha) } - 2 s \Bigg]- 2 m_1 m_2 (5 + 2
t + 5 t^2) \nnb \\
\ar 6 (1-t^2) m_q \Big[-(1-\alpha) m_1 - \alpha m_2 \Big]\Bigg\}~, \\ \nnb \\
\label{edhb13}
\rho_{12}^{(2)} \es {\sqrt{3} \over 128 \pi^4}
\int_{\alpha_{min}}^{\alpha_{max}}
{d\alpha \over \alpha^2} \int_{\beta_{min}}^{1-\alpha} {d\beta \over \beta^2}
(m_1^2 \beta + m_2^2 \alpha - s \alpha \beta)^2 (-2 + t + t^2) (\beta m_1 -
\alpha m_2) \nnb \\
\ar {m_q \qq \over 16 \sqrt{3} \pi^2} \int_{\alpha_{min}}^{\alpha_{max}}
d\alpha (-2+t+t^2) \Big[(1-\alpha) m_1 - \alpha m_2 \Big]~, \\ \nnb \\
\label{edhb14}
\rho_{22}^{(2)} \es {1 \over 256 \pi^4}
\int_{\alpha_{min}}^{\alpha_{max}} {d\alpha \over \alpha}
\int_{\beta_{min}}^{1-\alpha} {d\beta \over \beta}
(m_1^2 \beta + m_2^2 \alpha - s \alpha \beta) \nnb \\
\cp \Bigg[ (m_2 \alpha + m_1 \beta) (-1+t) (1+5 t) {m_1^2 \beta + m_2^2 \alpha -
s \alpha \beta \over \alpha \beta} - 6 (5 + 2 t + 5 t^2) m_1 m_2 m_q\Bigg]
\nnb \\
\ar {\qq \over 192 \pi^2} \int_{\alpha_{min}}^{\alpha_{max}} d\alpha
\Bigg\{ (1-\alpha) \alpha (-1+t) (13 + 11 t) \nnb \\
\cp \Bigg[ 3 m_0^2 +  {4 [m_1^2 (1-\alpha) + m_2^2 \alpha - s
\alpha (1-\alpha)]\over \alpha (1-\alpha) } - 2 s \Bigg] - 
6 (5 + 2 t + 5 t^2 ) m_1 m_2  \nnb \\
\ar 2 (1-t) (1+ 5t) m_q \Big[ - (1-\alpha) m_1 - \alpha m_2 \Big] \Bigg\}~,
\eea 
where,
\begin{eqnarray}
 \beta_{min}&=&\frac{\alpha m_2^2}{s\alpha-m_1^2},\nnb\\
\alpha_{min}&=&\frac{1}{2s}\Big[s+m_1^2-m_2^2-\sqrt{(s+m_1^2-m_2^2)^2-4m_1^2s}\Big],\nnb\\
\alpha_{max}&=&\frac{1}{2s}\Big[s+m_1^2-m_2^2+\sqrt{(s+m_1^2-m_2^2)^2-4m_1^2s}\Big].
\end{eqnarray}
Performing Borel transformation with respect to the variable $-q^2$ and
assuming quark--hadron duality we get,
\bea
\label{edhb15}
\Pi_{ij}^{(\alpha)} = \int_{(m_1+m_2)^2}^{s_0} \rho_{ij}^{(\alpha)}
e^{-s/M^2} ds~.
\eea
Substituting these expressions into Eq. (\ref{edhb03}), we obtain the
expression for the mixing angle in  the framework of the QCD sum rules
method.

Now we are ready to perform numerical calculations. For the numerical values of
the input parameters we use  $\qq (1~GeV) = -(246^{+28}_{-19}~MeV)^3$
\cite{Rdhb12},
$\bar{s}s = 0.8 \qq$, $m_0^2 = (0.8 \pm 0.2)~GeV^2$. For the masses of the
heavy quarks we use their $\overline{\rm MS}$ masses, which are given as
$\bar{m}_b(\bar{m}_b) = 4.16 \pm 0.03~GeV$, $\bar{m}_c(\bar{m}_c) = 1.28 \pm
0.03~GeV$ \cite{Rdhb13}, and $m_s (2~GeV) = 102 \pm 8~MeV$  \cite{Rdhb14}.
The expressions of the invariant functions contain three auxiliary
parameters, namely the Borel parameter  $M^2$, continuum threshold $s_0$
and an arbitrary parameter $t$. For the working regions of the
continuum threshold and Borel parameter we use the recent results obtained
from  analysis of the mass and residues of the doubly heavy baryons, i.e.,
$s_0=(45$--$56)~GeV^2$, and $6~GeV^2 \le M^2 \le 16~GeV^2$ \cite{Rdhb11}.
In the present study, the working regions of the parameter $t$ are also taken
to be $-0.72 \le \cos\theta \le -0.44$ and $0.44 \le \cos\theta \le
0.72$, where $t=\tan\theta$ (for details see \cite{Rdhb11}).

Considering these working regions for  auxiliary parameters, we obtain
$\varphi_{\Xi_{bc}} =  16^0 \pm 5^0$ for the $\Xi_{bc}$--$\Xi_{bc}^\prime$
case and $\varphi_{\Omega_{bc}} =  18^0 \pm 6^0$ corresponds to the
$\Omega_{bc}$--$\Omega_{bc}^\prime$ mixing. These results have been obtained for the
$\rlap/{q}$ structure. Very close results are also obtained using the $I$
structure.
The same mixing angles are also evaluated in \cite{Rdhb04} within the
non--relativistic quark model to have the values,
$\varphi_{\Xi_{bc}} = 25.5^0$ and $\varphi_{\Omega_{bc}} =
25.9^0$.
Comparing our results with these values, we see that
the predictions of the QCD sum rules are slightly smaller compared to that
of the non--relativistic quark model.

It should be noted here that, the consequence of mixing can considerably change
the results of semileptonic and electromagnetic decays of heavy baryons 
firstly pointed out in \cite{Rdhb02}.

In summary, we calculated the mixing angles between the doubly heavy 
$\Xi_{bc}$--$\Xi_{bc}^\prime$ and $\Omega_{bc}$--$\Omega_{bc}^\prime$ baryons using
the QCD sum rules method, and obtained that the mixing angles are quite large.
A comparison of our results on the mixing angles with the predictions of teh
non--relativistic quark model is also presented.

\end{document}